\documentclass[review]{article}
\usepackage[top=0.85in,footskip=0.75in,marginparwidth=2in]{geometry}
\usepackage{authblk}

\pdfoutput=1
\usepackage[utf8]{inputenc}

\usepackage{ragged2e}
\usepackage{cite}

\usepackage{nameref,hyperref}

\usepackage{microtype}
\DisableLigatures[f]{encoding = *, family = * }

\raggedright
\usepackage{changepage}

\usepackage[aboveskip=1pt,labelfont=bf,labelsep=period,singlelinecheck=off]{caption}

\makeatletter
\renewcommand{\@biblabel}[1]{\quad#1.}
\makeatother

\usepackage{lastpage,fancyhdr,graphicx}
\usepackage{epstopdf}
\usepackage{color}

\usepackage{amssymb}
\usepackage[T1]{fontenc}
\usepackage[utf8]{inputenc}
\usepackage{amsmath}
\usepackage{verbatim}
\usepackage[colorinlistoftodos]{todonotes}
\usepackage{multirow}
\usepackage{color}

\definecolor{Gray}{gray}{.25}

\usepackage{graphicx}

\usepackage{sidecap}

\usepackage{wrapfig}
\usepackage[pscoord]{eso-pic}
\usepackage[fulladjust]{marginnote}
\reversemarginpar
\title{Comparison of Baseline Correction Methods for LIBS Technique for Total Carbon Quantification in Brazilian Soils}
\author[1]{Marco Aur\'elio de Menezes Franco\thanks{marco.franco@usp.br}}
\author[2]{D\'ebora Marcondes Bastos Pereira Milori}
\author[2]{Paulino Ribeiro Villas-Boas}
\affil[1]{Institute of Physics, University of S\~ao Paulo, S\~ao Paulo, SP, Brazil}
\affil[2]{Embrapa Instrumentation, S\~ao Carlos, SP, Brazil}

\date{}
\begin{document}
\maketitle
% \vspace*{0.35in}

% % title goes here:
% \begin{flushleft}
% {\Large
% \textbf\newline{Comparison of Algorithms for Baseline Correction of LIBS Spectra for Quantifying Total Carbon in Brazilian Soils}
% }
% \newline
% % authors go here:
% \\
% Marco Aur\'elio de Menezes Franco\textsuperscript{1,*},
% D\'ebora Marcondes Bastos Pereira Milori\textsuperscript{2},
% Paulino Ribeiro Villas Boas\textsuperscript{2},
% \\
% \bigskip
% \bf{1} Institute of Physics, University of S\~ao Paulo, S\~ao Paulo, SP, Brazil
% \\
% \bf{2} Embrapa Instrumentation, S\~ao Carlos, SP, Brazil
% \\
% \bigskip
% * marco.franco@usp.br

% \end{flushleft}

\section*{Abstract}
\justifying
LIBS (Laser-Induced Breakdown Spectroscopy) is a promising and versatile technique for conducting multi-element analysis. It offers the advantage of rapid analysis, typically taking less than a minute, and requires minimal sample preparation. However, despite recent advancements in elemental quantification using LIBS, there are still challenges associated with the baseline produced by background radiation. This radiation introduces non-linear interference to the emission lines, thereby necessitating proper baseline correction for the creation of calibration models to quantify elements based on LIBS spectra. In this study, we conducted a comparative analysis of various filters and methods to address the issue of random noise removal and baseline correction in LIBS spectra, specifically focusing on the quantification of total carbon in soil samples. We systematically tested all combinations of filters and methods, optimizing their parameters to establish the strongest correlation between the corrected spectra and the carbon content in a training sample set. Subsequently, we compared the performance of these combinations, utilizing the optimized parameters, against a separate test sample set. Through rigorous evaluation, we identified that the combination of the Savitzky-Golay filter and the 4S Peak Filling method yielded the most effective baseline correction. This particular combination demonstrated a Pearson's correlation coefficient of 0.93, accompanied by a root mean square error of 0.21. Notably, this result surpassed the performance achieved by employing a linear regression model solely based on the carbon emission line at 193.04 nm. The linear regression model yielded a correlation of 0.91 but with a root mean square error of 0.26. The proposed procedure in this study presents a novel approach for baseline correction in LIBS spectra. Moreover, it opens up new possibilities for the development of multivariate methods utilizing a specific spectral range.

% now start line numbers
%\linenumbers

% the * after section prevents numbering
\section*{Introduction}

\begin{figure}[!ht]
 \centering
 \includegraphics[width=\linewidth]{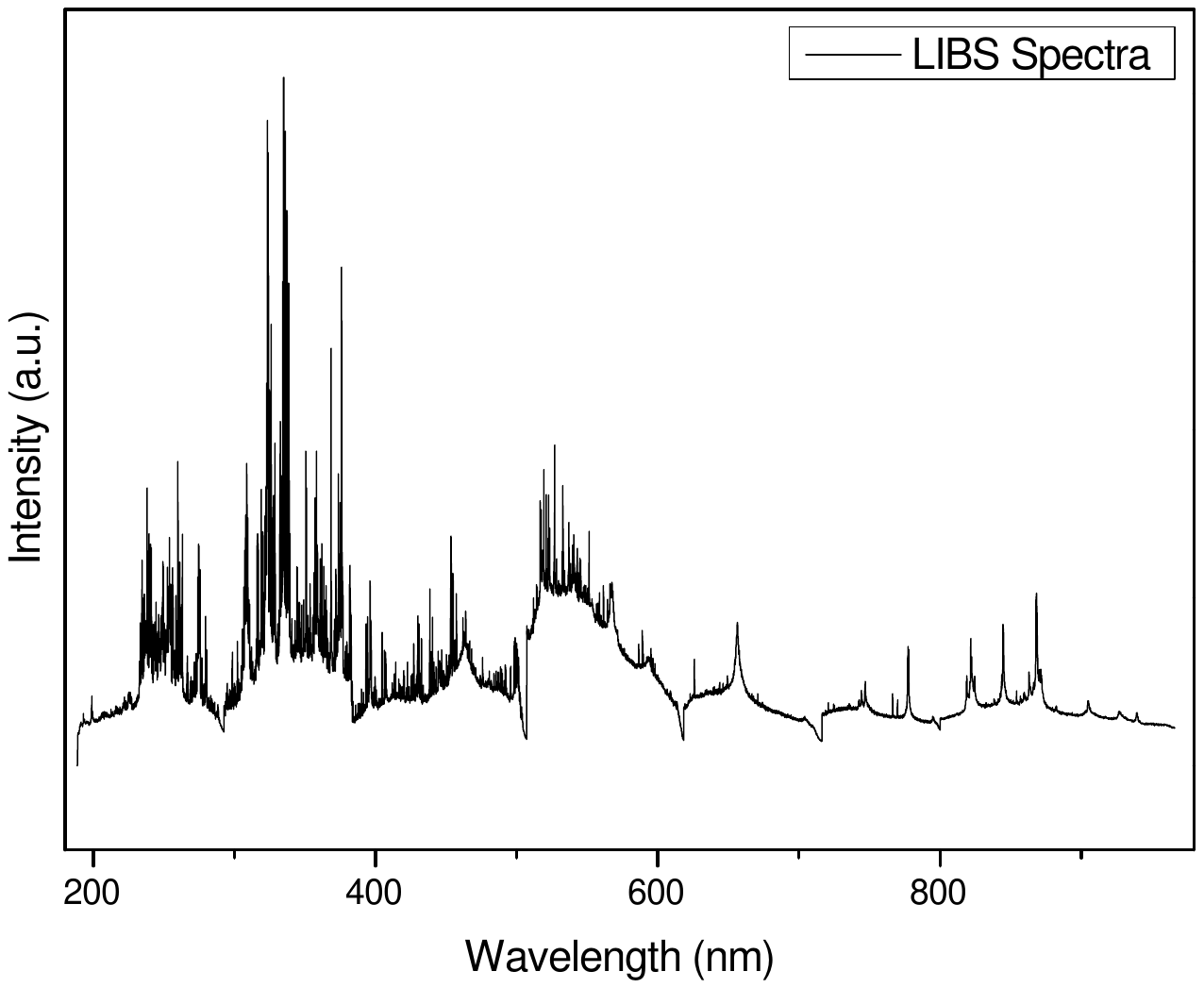}
 \caption{An example of an uncorrected LIBS spectrum is shown, displaying a noticeable irregular baseline. The spectrum is divided into seven distinct regions, each corresponding to a different spectrometer.}
 \label{fig:fig0}
\end{figure}

Laser-induced breakdown spectroscopy (LIBS) is a multi-element analytical technique widely employed in the last years~\cite{cremers2006laser, santos2015performance, ferreira2014novel, rusak1997fundamentals, hahn2012laser, stuart1996nanosecond} and has great potential to be applied in field~\cite{harmon2005laser, cunat2005portable, ferretti2007situ}. A LIBS measurement usually takes less than a minute and can be remotely applied in 
hazardous areas or of difficult access, such as Mars, deep sea water, and radioactive or toxic places~\cite{hoge1983airborne, salle2004laser, yun2002laser, buckley2000implementation}. LIBS also satisfies the precept of ``green chemistry'', in which no toxic residue is generated during the analytical process since the technique does not require chemical reagents~\cite{cremers2006laser, senesi2014laser}. LIBS is also considered a promising technique for many
applications, particularly for soil analysis ranging from the quantification of carbon~\cite{nicolodelli2014quantification}, nutrients
~\cite{hussain2007measurement} to contaminants~\cite{pandhija2010contaminant}. Nonetheless, the technique suffers from a low reproducibility and repeatability
compared to other analytical techniques (e.g.\ inductively coupled plasma - ICP), mainly due to matrix effects~\cite{mohamed2007study, eppler1996matrix, segnini2014physical}. Such effects comprise the radiation-matter interaction, composition, aggregation state of samples,
homogeneity, laser pulse energy density, and optical alignment. Each LIBS measurement, therefore, results in a unique plasma formation, whose line emissions and background radiation depend on. To overcome such difficulties, hundreds of LIBS measurements are generally performed and averaged for each sample, and the delay time between the laser shot and spectral acquisition is adjusted for the best signal-to-noise ratio and for attenuating the background radiation. Even with the adjustment of delay time, the background radiation is still present in the spectra, which corresponds to the spectral baseline as in Figure~\ref{fig:fig0}, and imposes difficulties for quantitative elemental analysis. \par

One commonly employed approach to address background radiation issues in LIBS is the baseline correction method of subtracting a straight line that connects the neighborhood of the emission line of interest \cite{dawson1993background}. However, this solution is limited to non-interfered emission lines where the neighborhood clearly belongs to the baseline without any other emission lines. In particular, carbon estimates from automatic algorithms for LIBS are still a challenge. This study aims to identify an optimal combination of random noise filters and baseline correction methods that minimize the influence of background radiation on calibration models for soil carbon quantification. We conducted a comparative analysis of three filters, namely Low Pass~\cite{antoniou2006digital}, Median~\cite{huang1979fast}, and Savitzky-Golay~\cite{savitzky1964smoothing}, to eliminate random noise. Additionally, we evaluated the performance of five baseline correction methods: Statistics-Sensitive Non-Linear Iterative Peak-Clipping (SNIP)\cite{ryan1988snip,morhavc2009algorithm}, Top Hat (TH)\cite{van1992fast,gil1996computing}, Median~\cite{gil1996computing}, Rolling Balls (RB)\cite{kneen1996algorithm}, and 4S Peak Filling (PF)\cite{liland20154s}. It is worth noting that these methods were originally developed for other photonic techniques such as X-ray, Raman, infrared, and atomic mass spectroscopy. \par

To identify the most effective combinations of filters and baseline correction methods, we optimized each combination using a training set of soil samples. The optimization aimed to achieve the highest correlation with a reference technique. Subsequently, all optimized combinations were evaluated using a separate validation set. Following the filtering and baseline correction steps, the corrected spectra served as input variables for partial least square regression. The resulting models were then subjected to 10-fold cross-validation to assess their performance. Pearson's correlation coefficient ($\rho$) and root mean square error (RMSE) were employed as evaluation metrics for each combination. Finally, the combinations were ranked based on their $\rho$ values.

\section{Materials and methods }

\subsection{Samples and reference measures}

In this study, we used a dataset consisting of 60 samples collected from farmlands across various regions of Brazil. These samples were previously characterized by the Agronomic Institute of Campinas, located in São Paulo State, Brazil, in a similar manner to previous studies~\cite{villas2016laser}. To prepare the samples for analysis, a series of steps were followed. Firstly, the samples were dried at $40^\circ$C. Subsequently, they were homogenized and ground using a mortar and pestle until the resulting particles were smaller than 0.15 mm. Following that, pelletization was carried out by applying an 8-ton press for approximately 30 seconds. To determine the reference carbon content of the samples, measurements were performed using a Perkin-Elmer 2400 CHNS/O analyzer, following the guidelines provided in the manufacturer's manual. \par

\subsection {LIBS spectroscopy}

\begin{figure}[!ht]
 \centering
 \includegraphics[width=\linewidth]{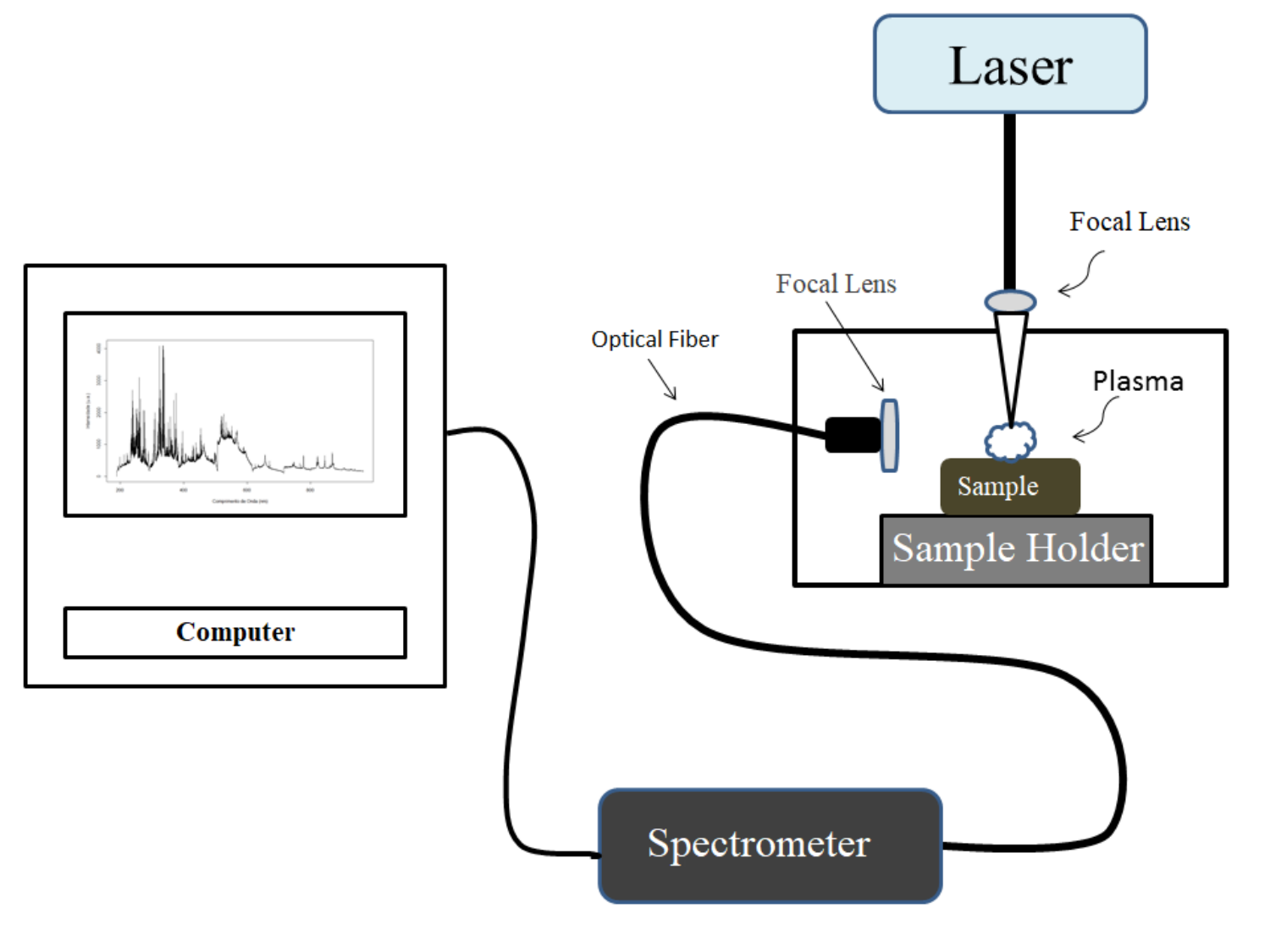}
 \caption{The experimental apparatus of the LIBS system consisted of several components, including a laser, an optical set, an ablation chamber, a set of spectrometers with seven optical fibers, and a computer.}
 \label{fig:libs} %Este comando você usa para dar um nome pra você chamar no seu texto. Pra chamar, use: \ref{nome}
\end{figure}

The LIBS measurements were conducted using an Ocean Optics LIBS2500+ spectrometer from Dunedin, USA. The spectrometer was equipped with a pulsed Nd: YAG Q-switched laser with a wavelength of $1064nm$, produced by Quantel (Big Sky Laser Ultra 50). The laser operated at a maximum energy of $50mW$ and a frequency of $10~Hz$, which was sufficient for plasma formation. The laser beam was focused onto the sample surface within an ablation chamber, and the emitted plasma radiation was collected by a system of optical fibers connected to seven spectrometers. The LIBS measurements were performed in atmospheric air, and the spectral range covered was from $189nm$ to $966nm$, with an optical resolution of $0.1nm$. The emitted spectra generated by the excited species were analyzed in the UV-VIS-IP regions. To enhance the precision of the results, a total of 60 spectra were captured for each soil sample, with each spectrum corresponding to a different location on the sample. The experimental setup of the LIBS system used for these measurements is depicted in Figure\ref{fig:libs}.\par

\subsection{Spectral processing}

\subsubsection{Tradicional baseline correction}

The traditional baseline correction method, commonly referred to as the three-point method~\cite{dawson1993background}, is a heuristic approach. This method involves correcting the baseline by subtracting the spectral region around the peak of interest using a straight line that connects the sides of the peak. By applying this correction, it becomes possible to obtain values for the peak height or peak area, which can then be used to generate models for uni or multivariable linear regression. The peak height represents the maximum intensity of the corrected peak, while the peak area corresponds to the integral of the peak or the area of a fitted distribution (such as Gaussian, Lorentzian, or Voigt). These parameters serve as important indicators for quantification and can be utilized in various analytical models.

\subsubsection{Noise suppression}

In order to perform automatic baseline correction, three signal smoothing filters were applied to the spectra. These filters include the Savitzky-Golay Filter (SGF), Low Pass Filter (LPF), and Median Filter. The SGF is a widely used filter that smooths the data, effectively enhancing the signal-to-noise ratio without significantly distorting the underlying signal. It achieves this by convolving successive subsets of adjacent data points with a polynomial using the method of linear least squares~\cite{savitzky1964smoothing}. The LPF, which has been improved by the research group, incorporates a Voigt profile~\cite{olivero1977empirical}. This profile selectively attenuates noises that exceed a certain cutoff frequency. The parameters of the convolution function are adjusted accordingly to achieve the desired noise reduction. Lastly, the Median Filter replaces each input element (i.e., noise) with the median value of the neighboring inputs. The window size, known as the standard of neighbors, moves across the entire spectrum~\cite{huang1979fast}. By applying these signal smoothing filters, the goal is to reduce noise and improve the quality of the spectra, thereby facilitating subsequent baseline correction procedures.

\subsubsection{Automatic methods for baseline correction}

For the purpose of baseline correction, several methods were first selected based on a general visual inspection. These methods include SNIP, TH, Median, RB, and FP~\cite{gibb2012maldiquant, liland2010optimal, morhavc2009algorithm, ryan1988snip, van1992fast, gil1991computing, liland20154s}. The SNIP method (Statistics-Sensitive Non-Linear Iterative Peak-Clipping) is an algorithm originally developed for atomic mass spectroscopy. It is designed to obtain the baseline in three stages, independently of the data acquisition system. This method employs a digital filter known as the "Low Statistics" filter, which smooths the spectrum based on a small number of pixels per window. TH (Top Hat) is a structured method that utilizes the Fourier transform of the spectrum intensity. It extracts fine details and small elements from the spectrum data that are more relevant than the neighboring noise. This is achieved through a sliding window, with the size of the window specified by the user. By applying these baseline correction methods, it is possible to enhance the accuracy and reliability of the spectra analysis, ensuring that the underlying signals are accurately captured while minimizing the impact of noise and interference.\par

The Median method for baseline correction is based on local average calculations. Each spectral point is replaced by the average value of its neighboring points. These averages are computed within local windows, which can be defined, for example, around peaks or within noise regions. The process begins by taking the average within a subset of data, determined by the user. Then, the subset is shifted by a window offset to create a new subset, and the average is recalculated. This iterative process is repeated multiple times throughout the dataset. Rolling Balls (RB) is a method originally developed for processing X-ray spectroscopy data. It utilizes the concept of a rolling sphere with a fixed radius that moves along the spectrum. The trajectory of the rolling sphere creates a smoothed line that is used to extract the baseline from the spectrum. The 4S Peak Filling method (PF) is an iterative approach for baseline suppression based on the averaging of local windows. It requires the second derivative of the spectrum for primary smoothing, assuming that the baseline should exhibit a monotonically decreasing trend. The method also takes into account parameters such as the average size of the data window, the number of windows used, and the number of iterations performed during the baseline suppression loop. Within each local window, the minimum value from the averages and iterations is selected as the new baseline. By applying these methods, it is possible to effectively correct the baseline of the spectra, eliminating unwanted variations and enhancing the accuracy of subsequent data analysis.\par

\subsubsection{Calibration model}

%The method of multivariate regression by Partial Least Squares (PLS) was used to construct a calibration model of the LIBS spectra for prediction of total carbon in the samples. This method was used because it reduces the dimension of the input variables and maximizing the correlation with the response variable~\cite{martins1996curso, cunha2007regressao}. With the established prediction model, was made the validation of the results obtained with those of the reference method, which gives the Pearson coefficients that measure the degree of correlation between two variables. They are represented by $\rho$ and assumes values between $-1$ and $1$, where $\rho\geq0.7$ indicates strong correlation, $0.3<\rho<0.7$ indicates moderate correlation and $\rho\leq0.3$ corresponds to a weak correlation between the results~\cite{lawrence1989concordance}. The most used value to quantify differences between estimated values by the model and the reference measures is the root mean square uncertainty (RMSE) of the correlation, which can be estimated by the equation::
%\begin{equation}
%RMSE=\sqrt{\frac{1}{n}\sum\limits_{i=1}^{n}{(y_{i}-\hat{y}_i)^{2}}},
%\label{eq6}
%\end{equation}
%where $\hat{y}_i$ is the value estimated by the model for the $i$-th sample and $n$ is the number of samples.

The Partial Least Squares (PLS) method, a multivariate regression technique, was employed to develop a calibration model using the LIBS spectra for predicting the total carbon content in the samples. PLS was chosen due to its ability to reduce the dimensionality of the input variables while maximizing the correlation with the response variable~\cite{martins1996curso, cunha2007regressao}. To validate the accuracy of the prediction model, the obtained results were compared with those obtained from the reference method. The degree of correlation between the two variables was measured using Pearson coefficients, denoted as $\rho$. The range of $\rho$ values is between -1 and 1, where $\rho\geq0.7$ indicates a strong correlation, $0.3<\rho<0.7$ indicates a moderate correlation, and $\rho\leq0.3$ corresponds to a weak correlation between the results~\cite{lawrence1989concordance}. To quantify the differences between the estimated values by the model and the reference measurements, the root mean square uncertainty (RMSE) of the correlation is commonly used. The RMSE can be estimated using the following equation:

\begin{equation}
RMSE=\sqrt{\frac{1}{n}\sum\limits_{i=1}^{n}{(y_{i}-\hat{y}_i)^{2}}},
\label{eq6}
\end{equation}

where $\hat{y}_i$ represents the value estimated by the model for the $i$-th sample, and $n$ denotes the total number of samples. The RMSE provides a measure of the average difference between the predicted and reference values, allowing for an assessment of the model's predictive performance. \par

\subsubsection{Analysis proceedings}

%The approach to the problem of determining the best baseline correction method was as follows: the set of corrected spectra were obtained and divided into two parts, $\frac{2}{3}$ of the total spectrum to build the PLS model and optimization signal filter parameters and line correction method, and the rest for test of the model. For the optimization was constructed a regression model with the corrected spectra and the carbon content measured by the reference method \emph{CHNS}. The model validation occurred through cross-validation, obtaining a value of $\rho$ and $\epsilon$, where $\rho$ corresponds to the Pearson coefficient, and $\epsilon$ represents the uncertainty of the model given in mass percentage of total carbon. This process was repeated iterated times, and the values of $\rho$ were saved, so that the end of the iterative process, $\bar{\rho}$ was calculated the the associated uncertainty. The greatest value of $\bar{\rho}$ with the lowest $\epsilon$ was used as a condition for determining the set of parameters. Thus, this set was applied to the second set of spectra for building a regression model that was validated by cross validation. 

The approach taken to determine the optimal baseline correction method was as follows: The set of corrected spectra was divided into two parts, with $\frac{2}{3}$ of the total spectra used to build the Partial Least Squares (PLS) model and optimize the parameters of the signal filters and baseline correction methods. The remaining $\frac{1}{3}$ of the spectra were used to test the model. To optimize the parameters, a regression model was constructed using the corrected spectra and the carbon content measured by the reference method (CHNS). Cross-validation was performed to validate the model, obtaining values for the Pearson coefficient ($\rho$) and the uncertainty ($\epsilon$) of the model, expressed as the mass percentage of total carbon. This process was iterated multiple times, and the values of $\rho$ were recorded. At the end of the iterative process, the average value of $\rho$ ($\bar{\rho}$) and its associated uncertainty were calculated. The set of parameters that yielded the highest $\bar{\rho}$ with the lowest $\epsilon$ was selected as the optimal condition. This set of parameters was then applied to the second set of spectra to build a regression model, which was further validated using cross-validation.

\section{Results}

\begin{figure}[!ht]
 \centering
 \includegraphics[width=\textwidth]{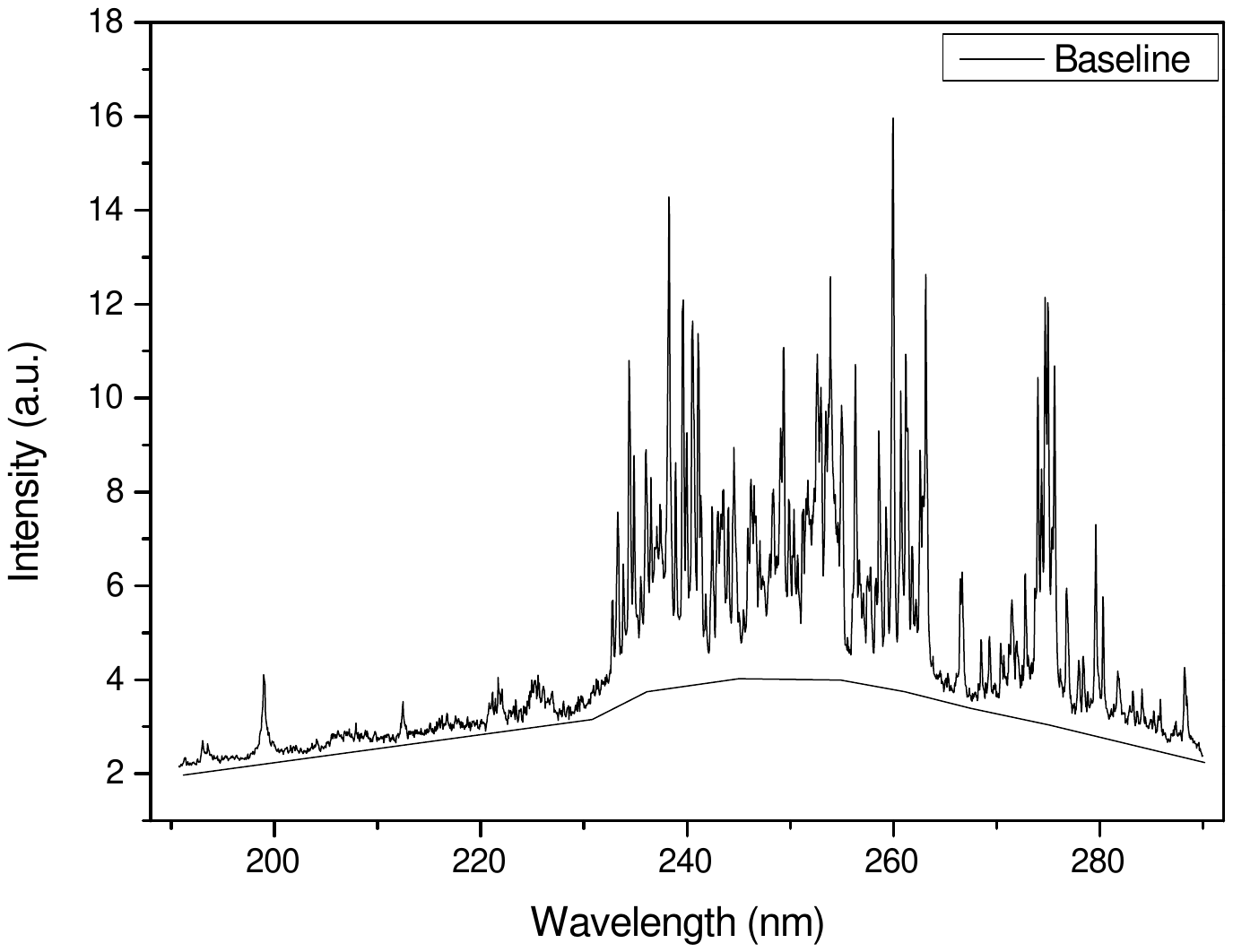}
 \caption{Average spectrum from a soil sample without baseline correction and a hypothetical baseline predicted for the wavelength range.}
 \label{fig:semcorrecao}
\end{figure}

The focus of the study was on the spectral range of $190-289nm$, which corresponds to the first spectrometer of our instrument. Within this range, there are two specific peaks of interest for the determination of total carbon in the soil, located at $193.04nm$ and $247.86~nm$. However, this particular spectral range poses significant challenges in determining the appropriate baseline due to the presence of multiple chemical species emitting radiation simultaneously and the strong presence of radiation at all wavelengths (referred to as white radiation), primarily caused by the thermal effect of the plasma. Figure~\ref{fig:semcorrecao} illustrates the average LIBS spectrum of and a hypothetical baseline that could potentially correct the spectral range. It is evident from the figure how the continuous background radiation significantly interferes with the spectral shape, resulting in a pronounced baseline, particularly in the region between $230$ and $265~nm$. It is worth noting that this spectral range exhibits numerous emissions arising from different ionization degrees of various chemical species, such as Fe, C, Al, and Si.

The traditional baseline correction method proves to be limited in its ability to correct the baseline in the spectral range surrounding the peak of interest at $193.04nm$. However, this method fails when applied to another peak located at $247nm$, as it is surrounded by numerous other emission lines, making it challenging to find a suitable lateral region for baseline correction. To address this limitation, a linear model was constructed using the peak area at $193.04~nm$ and carbon concentrations. The model achieved a correlation coefficient of $\bar{\rho}_{training} = 0.92$ with an uncertainty of $\epsilon_{training} = 0.24$ for the training set, and $\bar{\rho}_{test} = 0.91$ with an uncertainty of $\epsilon_{test} = 0.26$ for the test set, where $\epsilon$ represents the uncertainty in the measurement of total carbon content. \par

To overcome the challenges posed by baseline correction, various combinations of baseline correction methods (PF, RB, TH, Median, and SNIP) and smoothing filters (Median, LPF, and SGF) were applied independently to the training and test sets. The optimized parameters for these combinations are presented in Table~\ref{tab_param}, while Table~\ref{tab_comp} displays the Pearson coefficients and their corresponding uncertainties for the cross-validation of reference measurements and the prediction model for total carbon content.\par

\begin{table}[!ht]
\caption{Set of optimized parameters through training set, applied to the test set, with respect to baseline correction methods and noise smoothing filters.} 
	
    \begin{center}
    	\begin{tabular*}{0.99\linewidth}{@{\extracolsep{\fill} } l l c c c c c} 		 \hline
    	Method & Filter & Filter Par. & $\lambda$ & $\beta$ & I & $\alpha$ \\ \hline
    	{SNIP}	& SGF & [O.P.,P.J.] = [8, 13] & -& - & 100 & - \\ 
	 	& $ LPF$ & $\sigma = 0.44$, $\gamma = 10.25$ & - & - &103 & - \\  
    	& Median & N= 1 & - & - & 14 & - \\    
    	{TH}
       & SGF & [O.P.,P.J.] = [8, 15] & -& 274 & - & - \\ 
	 	& $ LPF$ & $\sigma = 0.20$, $\gamma = 4.6$  & - & 276 & - & - \\  
    	& Median & N= 2 & - & 14 & - & - \\
    	Median
        & SGF & [O.P.,P.J.]= [1, 13] & -& 260 & - & - \\ 
	 	& $ LPF$ & $\sigma = 1.22$, $\gamma = 16.22$  & - & 277 & - & - \\  
    	& Median & N= 1 & - & 14 & - & - \\
    	{RB}
         & SGF & [O.P.,P.J.] = [6, 13] & -& [wn,ws] = [340, 30] & - & - \\ 
	 	& $ LPF$ & $\sigma = 0.43$, $\gamma = 11$  & - & [wn,ws] = [374,47] & - & - \\  
    	& Median & N= 3 & - & [wn,ws] = [340,30] & - & - \\
    	{PF}
        & SGF & [O.P.,P.J.] = [8, 13] & 3& 10 & 9 & 70 \\ 
	 	& $ LPF$ & $\sigma = 0.20$, $\gamma = 4.6$  & 3 & 10 & 9 & 70 \\  
    	& Median & N= 2 & 3 & 10 & 9 & 70 \\ \hline
    	\end{tabular*}
    \end{center}
    O.P. = polynomial order \\
    P.J. = Window points for smoothing \\
    I = Number of iterations \\
    N = Number of points for moving window Median filter \\
    $\sigma$, $\gamma$ = Setting parameters of Voigh profile curve of the LPF \\
    $\lambda$ = penalty parameter to the second derivative to primary smoothing, according to \cite{liland20154s}\\
    $\alpha$ = Number of areas where spectrum is divided \\
    $\beta$ = window number of points \\
    wn = local window points to minimize/maximize \\
    ws = local window points for smoothing 
     
    \label{tab_param}
\end{table}

\begin{table}[!ht]
\caption{Pearson Coefficient ($\bar{\rho}$) and RMSE($\epsilon$) to validate the model \emph{PLS} with relation to the various noise filter combinations and baseline correction method for training and test sets.} 
	
    \begin{center}
    	\begin{tabular*}{0.99\linewidth}{@{\extracolsep{\fill} }l l c c c c} 		 \hline
    	Method & Filter & $\bar{\rho}_{treining}$ & $\epsilon_{treining}$ & $\bar{\rho}_{test}$ & $\epsilon_{test}$ \\ \hline
    	{SNIP}	& SGF &0,96 & 0.17 &0.92 & 0.21 \\ 
	 	& $ LPF$ & 0.95& 0.17& 0.91 & 0.23 \\  
    	& Median &0.95 &0.19& 0.93 & 0.21 \\    
    	{TH}
        & SGF & 0.96& 0.17& 0.93 & 0.20 \\
	 	& $ LPF$ &0.96 & 0.18& 0.92 & 0.21   \\  
     	& Median & 0.96 & 0.19& 0.88 & 0.26  \\ 
    	{Median}
        & SGF &0.97& 0.16& 0.92 & 0.21 \\ 
		& $ LPF$ &0.96 & 0.18 & 0.92 & 0.22 \\  
     	& Median &0.94 & 0.22& 0.90 & 0.24 \\ 
    	{RB}
        & SGF &0.95 & 0.19& 0,93 & 0,20\\ 
	 	& $ LPF$ &0.95 & 0.19& 0.92 & 0.21   \\  
     	& Median &0.95 & 0.19& 0.92 & 0.21  \\ 
    	{PF}
        & SGF &0.96 &0.17&0.93 & 0.21 \\ 
	 	& $ LPF$ &0.95 & 0.19& 0.91 & 0.22   \\  
     	& Median &0.95 &0.20& 0.91 & 0.23 \\
        {Traditional}
        & &0.92 & 0.24&0.91 & 0.26  \\ 
	 	 \hline
    	\end{tabular*}
    \end{center}
    \label{tab_comp}
\end{table}

The SNIP method proved to be effective in correcting the spectra, preserving the shape of the peaks of interest without significant distortion. When combined with the SGF filter, it yielded notable results, particularly in complicated spectral ranges such as $230$ to $265~nm$, where a smooth baseline was achieved as anticipated. However, when SNIP was applied with the Median and LPF filters, the baseline was overestimated, leading to the deformation of peaks and the overall spectrum shape. Consequently, these data treatments are unsuitable for the analysis of carbon and other chemical species. Among the combination of methods and filters tested, the SNIP-SGF combination yielded the best results. It achieved a correlation coefficient of $\bar{\rho}_{training} = 0.96$ with an uncertainty of $\epsilon_{training} = 0.17$ for the training set, and $\bar{\rho}_{test} = 0.92$ with an uncertainty of $\epsilon_{test} = 0.21$ for the test set. These findings demonstrate the effectiveness of the SNIP-SGF combination in accurately analyzing and predicting carbon content in the samples.\par 

\begin{figure}[!ht]
 \centering
 \includegraphics[width=\linewidth]{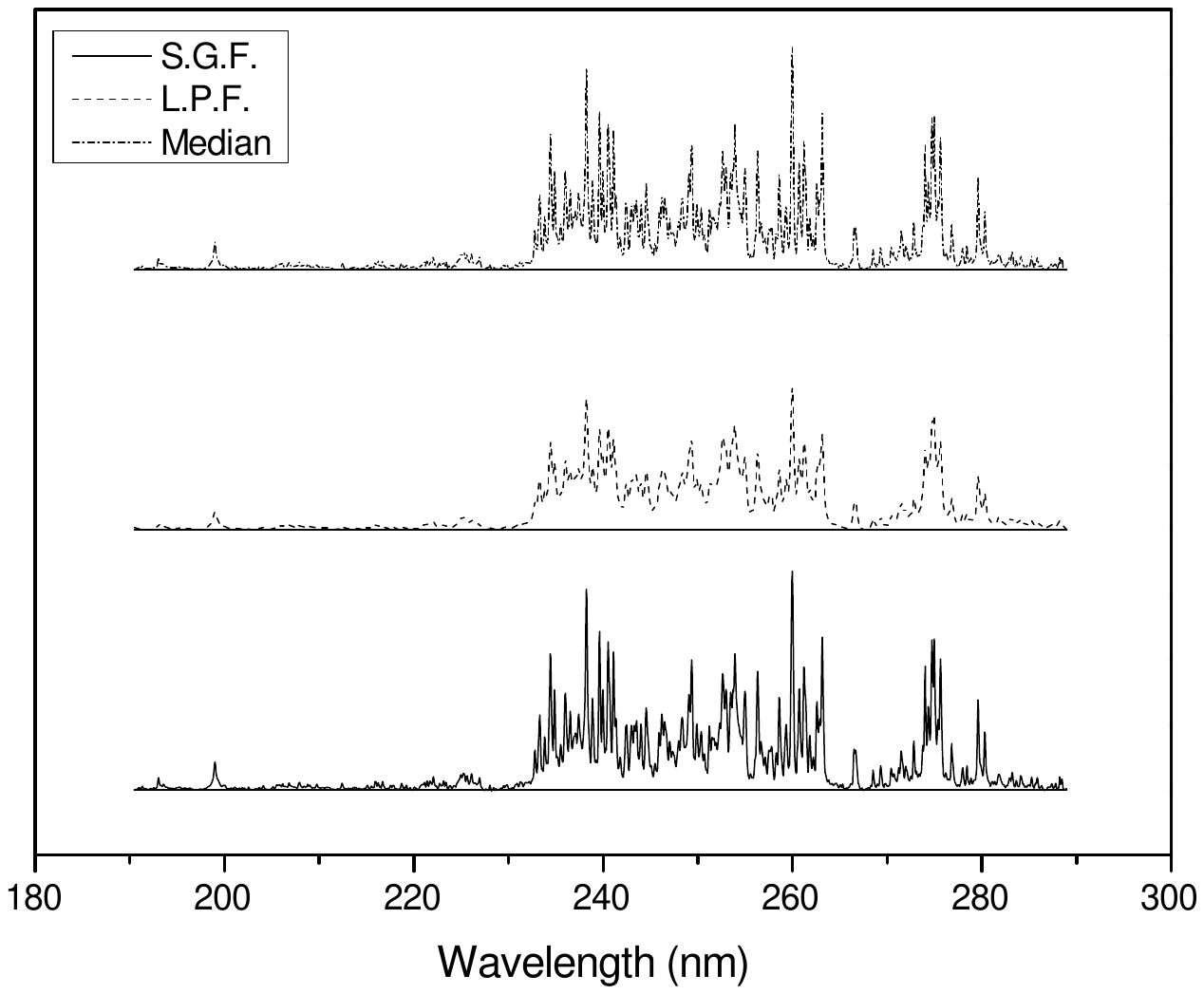}
 \caption{Spectra corrected by the 4S Peak Filling method for three different smoothing filters.}
 \label{fig:FP}
\end{figure}

The spectra corrected using combinations of the three smoothing filters with the TH method exhibited difficulties in establishing a useful baseline due to the inclusion of unexpected spectrum points in the baseline determination process. As a result, all calculated baselines were overestimated, even in cases where there was a high correlation between the calibration and reference models. Similar issues were observed when applying the previously discussed procedure to the Median method. In all method-filter combinations, the resulting spectra displayed numerous peaks with overestimated baselines, leading to negative values for the relative intensity of peaks. This clearly demonstrates that the combinations of the Median method with the three types of smoothing filters are also unsuitable for the treatment of LIBS spectra.\par 

The combination of RB with the three smoothing filters yielded satisfactory results for baseline correction, particularly in spectral ranges with fewer emissions. However, challenges were encountered in accurately estimating the baseline in transition regions where there is a mix of low and high emission points. Despite this limitation, the method demonstrated good correlation results, with $\bar{\rho}_{training}=0.95$ and $\epsilon_{training} = 0.19$ for the training set, and $\bar{\rho}_{test}= 0.93$ and $\epsilon_{test} = 0.20$ for the test set when RB was combined with SGF.\par

The PF method demonstrated superior performance in correcting LIBS spectra, particularly when combined with the SGF smoothing filter. This combination produced well-defined peaks and effectively corrected all spectral regions without compromising important spectral information required for accurate carbon prediction. The model constructed using the training set achieved impressive results, with $\bar{\rho}_{training}=0.96$ and $\epsilon_{training} = 0.17$. Similarly, the test set yielded favorable results, with $\bar{\rho}_{test}=0.93$ and $\epsilon_{test} = 0.21$. Additionally, the combination of \emph{4S Peak Filling} with the Median filter also delivered significant outcomes, surpassing the performance of LPF, which led to distorted spectra due to excessive smoothing. Figure~\ref{fig:FP} shows the corrected spectra of a sample for each PF/smoothing filter combination. \par

\section{Discussions}

\begin{figure}[!ht]
 \centering
 \includegraphics[width=\linewidth]{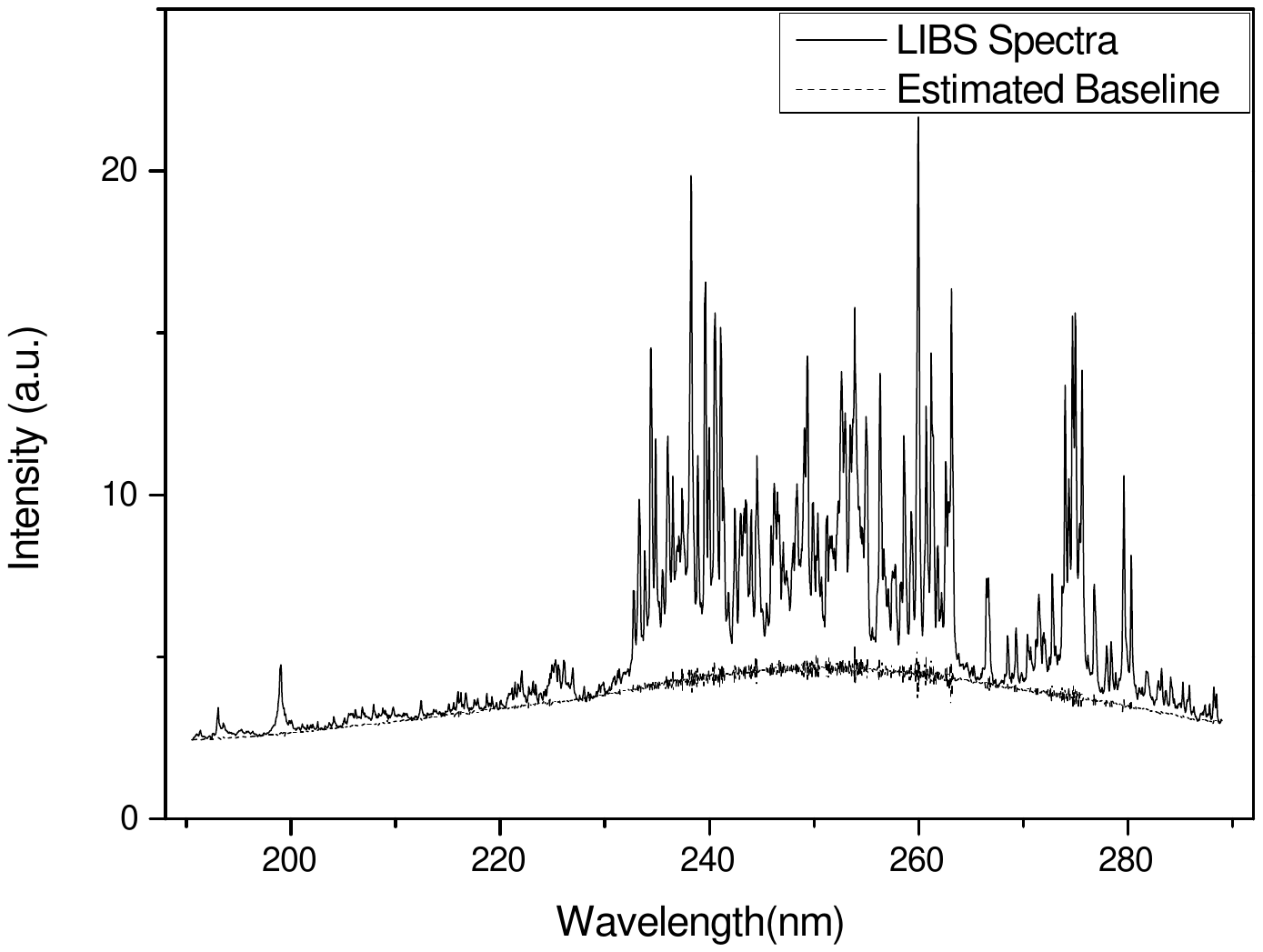}
 \caption{Baseline estimated by the 4S Peak Filling method for the spectral region between 193.04 nm and 247.86 nm.}
 \label{fig:baseline}
\end{figure}

The traditional baseline correction method proves effective when dealing with isolated peaks that have clear lateral regions, allowing for the construction of a straight line for correction. However, when it comes to the carbon peak at 193.04~nm, interference from aluminum complicates the correction process. Nevertheless, the method successfully resolves these interferences through bilorentzian deconvolution, enabling the calculation of the carbon peak area and the construction of a multivariate model using this value. Unfortunately, this method is only applicable to sparsely populated regions of emission lines, which is uncommon in LIBS spectra. Consequently, the traditional method's utility is greatly restricted in such cases.\par

The application of smoothing filters played a crucial role in shaping the peaks of different chemical species and removing noise, resulting in improved accuracy in total carbon quantification. Among the filters, the Savitzky-Golay Filter (SGF) proved particularly effective in modeling the spectra for all methods. This filter works by fitting a polynomial function of degree $n$ to the spectrum and its various spectral regions, even in complex regions such as between 230 and 265 nm. The black spectrum shown in Figure~\ref{fig:FP} represents the PF-SGF combination, demonstrating the powerful corrective capabilities of this technique. In the problematic region mentioned, a smooth baseline was successfully achieved, as shown in Figure~\ref{fig:baseline}.\par 

\section{Conclusions}

This study aimed to compare different baseline correction methods for the correction of LIBS spectra, as there is a lack of effective alternatives in the literature. The SNIP, Top Hat, Median, 4S Peak Filling, and Rolling Balls methods were evaluated in combination with the LPF, SGF, and Median noise filters, and compared to the traditional baseline correction method. The goal was to find the most effective data processing approach for LIBS spectra that maximizes carbon quantification. For each method combined with the filters, as well as for the traditional method, PLS models were constructed and evaluated using cross-validation. The performance of each method was optimized to improve the correlation between LIBS spectra and carbon concentration in the samples. Some filter combinations distorted the spectrum and compromised peak definition, making it impossible to quantify chemical elements. However, others significantly improved the spectrum's format, providing a clearer understanding of previously unclear spectral regions. The smoothing filters proved valuable in identifying and quantifying important peaks related to different chemical species, particularly carbon. \par

Notably, the combination of the 4S Peak Filling method with the Savitzky-Golay polynomial noise filter yielded a Pearson coefficient of $\bar{\rho_{test}} = 0.93$ and RMSE of $\epsilon_{test} = 0.21$ for the test set. This result was significantly better than that of the traditional baseline correction method, which achieved a Pearson coefficient of $\bar{\rho_{test}} = 0.91$ and RMSE of $\epsilon_{test} = 0.26$ when using the 193.04 nm peak. This demonstrates that the baselines of LIBS spectra can be effectively corrected using the 4S Peak Filling method in combination with the Savitzky-Golay noise filter. Consequently, the proposed data processing methodology in this study offers significantly improved robustness compared to the traditional method. Furthermore, the LIBS technique shows promise for the determination of total carbon in soils and potentially for quantifying other elements in complex matrices such as soil. The methodology proposed here can also be extended to the analysis of other elements.\par

\section{Acknowledgements}

This work has been supported by the Embrapa Instrumentation and Brazilian Research Council CAPES.

%{\small
%\bibliographystyle{elsarticle}
%\bibliography{ref}
%}
%\end{document}

%\nolinenumbers

%This is where your bibliography is generated. Make sure that your .bib file is actually called library.bib
\bibliography{ref}

%This defines the bibliographies style. Search online for a list of available styles.
\bibliographystyle{unsrt}

\end{document}